\documentclass[twocolumn,showpacs,preprintnumbers,amsmath,amssymb]{revtex4-1}
\usepackage{bm}
\usepackage{mathrsfs}
\usepackage{subfigure}
\usepackage{graphicx}
\usepackage{epstopdf}
\usepackage{color}

\begin{document}
\preprint{HUPD-1212}
\title{
Regularization parameter independent analysis in
Nambu--Jona-Lasinio model
}

\author{T.~Inagaki}
\affiliation{
Information Media Center, Hiroshima University,
Higashi-Hiroshima, Hiroshima
739-8521, Japan}
\author{D.~Kimura}
\affiliation{
Faculty of Engineering, Kinki University, Higashi-Hiroshima,
Hiroshima, 739-2116, Japan
}
\author{H.~Kohyama}
\affiliation{
Department of Physics, Kyungpook National University,
Daegu 702-701, Korea
}
\author{A.~Kvinikhidze}
\affiliation{
A. Razmadze Mathematical Institute of Georgian Academy of Sciences,\\
M. Alexidze Str. 1, 380093 Tbilisi, Georgia
}

\date{\today}

\begin{abstract}
Nambu--Jona-Lasinio model used to investigate low energy phenomena 
is nonrenormalizable, therefore the results depend on the regularization
parameter in general. A possibility of the finite in four-dimensional
limit and even the regularization parameter (this is dimension in
the dimensional regularization scheme) independent  analysis is shown
in the leading order of the $1/N_c$expansion.
\end{abstract}

\maketitle

\section{INTRODUCTION}
Nambu--Jona-Lasinio (NJL) model~\cite{NJL} is one of the
most popular QCD motivated effective theories used 
to understand non-perturbative  low energy phenomena of
strong interactions~%
\cite{Vogl:1991qt,Klevansky:1992qe,Hatsuda:1994pi}.

Its Lagrangian contains a four-fermion interaction,
an operator whose dimension exceeds the space-time dimension,
$D$, for $D>2$. It is known that the four-fermion interaction
model is renormalizable  in the $1/N_c$ expansion scheme for
$2<D<4$ and the model possesses an ultraviolet-stable point~%
\cite{Eguchi:1977kh, Shizuya:1979bv, Rosenstein:1988pt}.
It is nonrenormalizable in four space-time dimensions, however,
the ultraviolet divergences remaining in the renormalized Green's 
function are logarithmic in the leading order of the $1/N_c$
expansion~\cite{Shizuya:1979bv}.

In order to construct an effective theory out of infinite number
of operators involved in the NJL model one usually picks up
operators which make a major contribution to targeted
phenomena. For example, scalar type four-fermion operators
are considered.

As the NJL model is not renormalizable in four space-time dimensions
some regularization methods are used to avoid divergences of loop
integrals thereby to obtain  finite
values of predicted physical quantities.

Three-momentum sharp cutoff regularization is widely used.  
Other regularization procedures are also studied in NJL
type  models, e.g., the smooth cutoff~\cite{Plant:1997jr,Hell:2008cc}, 
the dimensional regularization (DR)~\cite{Krewald:1991tz,
Inagaki:1994ec, Jafarov:2004jw, Inagaki:2007dq, Fujihara:2008ae,
Inagaki:2010nb,Inagaki:2011uj,Inagaki:2012re}, 
the Pauli-Villars~\cite{Osipov:2004mn}  
and the Fock-Schwinger proper-time regularization~\cite{Inagaki:1997nv,%
Inagaki:2003yi}. The regularization parameter 
dependence is discussed in the three-momentum sharp cutoff 
and in the DR schemes
in~\cite{Fujihara:2008ae,Inagaki:2010nb,Inagaki:2011uj,Inagaki:2012re}.

In this paper we show that for some quantities the NJL model behaves
well in the ultraviolet limit. Moreover, the predicted quantities
have reasonable values in the 4D limit; they are close to those obtained
in the DR scheme where the dimension $D < 4$ is a regularisation
parameter.

The paper is organized as follows.  In Sec.~\ref{njl_model},
the three-flavor NJL model introduced. Then we briefly review 
the dimensional regularization. In Sec.~\ref{sec_phys},
we calculate meson masses, their decay constants, etc., 
in the leading order of the $1/N_c$ expansion using DR.
In Sec.~\ref{sec_strategy}, the
regularization parameter dependence is discussed.
In Sec.~\ref{sec_4D}, the four dimensional limit is considered. 
The order of divergences is evaluated for $n$ point
Green  functions in the leading order of the $1/N_c$ expansion. 
Finite values are obtained for physical 
observables. We also derive analytic relationships between
observables. In Sec.~\ref{result}, we phenomenologically fix
the model parameters and numerically evaluate the physical
quantities. Some concluding remarks are given in Sec.~%
\ref{conclusion}.

\section{NJL MODEL}
\label{njl_model}
\subsection{NJL model}
The three-flavor NJL model including Kobayashi-Maskawa-'t Hooft
term~\cite{Kobayashi:1970ji,'tHooft:1976fv} is given as,
\begin{equation}
 \mathcal{L}_{\mathrm{NJL}} = \sum_{i,j}
 \bar{q}_i\left( i \partial\!\!\!/
  - \hat{m}\right)_{ij}q_j + \mathcal{L}_4 + \mathcal{L}_6 ,
\label{LNJL}
\end{equation}
where
\begin{align}
 \mathcal{L}_4 &= G \sum_{a=0}^8 \Biggl[
  \Bigl( \sum_{i,j} \bar{q}_i\lambda_a q_j \Bigr)^2
  + \Bigl( \sum_{i,j} \bar{q}_i\,i \gamma_5 \lambda_a q_j \Bigr)^2
  \Biggr] ,
\label{L_4} \\
 \mathcal{L}_6 &= -K \left[ \det\bar{q}_i (1-\gamma_5) q_j 
 +\text{H.c.\ } \right] ,
\label{L_6}
\end{align}
the subscripts $i,j$ are the flavor indices, $i,j=u,d,s$,
and $\hat{m}$ denotes the current quark mass matrix, 
$\hat{m}=\mbox{diag}(m_u,m_d,m_s)$. Below we consider the
${\rm SU(2)}$  isospin symmetric case, $m_u=m_d$, for
simplicity. $\lambda_a$ are the Gell-Mann matrices in the
flavor space, $G$ and $K$ represent the effective coupling
constants for four- and six-fermion interaction, respectively.
$G$ and $K$ have negative mass dimensions, $-2$
and $-5$ respectively, so the model is nonrenormalizable in four
space-time dimensions. The determinant in $\mathcal{L}_6$ 
concerns the matrix elements labeled by the flavor indices.
We suppose the order of the coupling constants to be
$G N_c \simeq O(1)$ and $K N_c^2 \simeq O(1)$, where $N_c$
is the number of colors.

The chiral condensates $\langle \bar{i}i \rangle$ generate the
constituent quark masses, $m^*_i$, inside mesons.  One solves
the gap equations to evaluate the constituent  quark masses. In
the leading order of $1/N_c$  expansion, the gap equations are
obtained as follows~%
\cite{Vogl:1991qt, Klevansky:1992qe, Hatsuda:1994pi},
\begin{align}
  m_i^{*} =
  m_i + 4 G (i {\rm tr} S^i) + 2 K (i {\rm tr} S^j)
  (i {\rm tr} S^k),
\label{gap_eq} 
\end{align}
with $i\not=j\not=k\not=i$. ${\rm tr}S^i$ represent the chiral
condensates, $i{\rm tr}S^i=-\langle \bar{i}i \rangle$
which are given by the trace of the quark propagator,
\begin{equation}
 i\, {\rm tr} S^i = -\int \frac{d^Dp}{i(2\pi)^D} {\rm tr} S^i(p)  
\label{trS} 
\end{equation}
where
\begin{equation}
 S^i(p) \equiv \frac{1}{p\!\!\!/-m_i^*+i \varepsilon}  
\end{equation}
and $D(\equiv 4- 2\epsilon)$ is the space-time dimensions for
internal quark fields. ``tr''
in the integral 
denotes the trace with respect to the spinor
and color indices.

\subsection{Dimensional regularization}
\label{subsec_DR}
The quark loop integral in Eq.~(\ref{trS}) is divergent in four
space-time dimensions. One has to regularize it to obtain a finite
result. The regularization dependence for the physical quantities 
are induced by this procedure. Using the 
DR we are going to take the four space-time dimensional limit.

In the DR we have
\begin{equation}
 i\, {\rm tr} S^i = \frac{N_c}{(2\pi)^{D/2}} 
 \Gamma\left( 1-\frac{D}2 \right) m_i^* (m_i^{*2})^{D/2-1} .
\label{trS2} 
\end{equation}
where the mass dimension is a function of $D$. 
The integral like Eq.~(\ref{trS2}), needs to be
multiplied by the 
mass scale parameter $M_0^{4-D}$ in order to correct the mass
dimension.
Note that in the previous studies~\cite{Inagaki:2010nb,Inagaki:2011uj,
Inagaki:2012re},
this parameter is called ``the renormalization scale''. However,
in this paper, we name it ``mass scale parameter'' to clearly
distinguish our treatment from the renormalization.
Then, the ``rescaled'' chiral condensates are given by
\begin{equation}
  \langle \bar{i}i \rangle_{\mathrm{rs}}
  = M_0^{4-D} \langle \bar{i}i \rangle.
\label{phi_r}
\end{equation}

It is also important to discuss the mass dimension of the coupling
constants $G$ and $K$. Substituting Eq.~(\ref{trS2}) into the gap
equations~(\ref{gap_eq}) tells us that the mass dimensions of $G$
and $K$ are $2-D$ and $3-2D$, respectively. Then the rescaled
couplings are evaluated as
\begin{equation}
  G_{\mathrm{rs}} = M_0^{D-4} G, \quad
  K_{\mathrm{rs}} = M_0^{2(D-4)}K .
\label{GKr}
\end{equation}
It is worth mentioning that $G \,{\rm tr}S$ and $K ({\rm tr}S)^2$
do not depend on $M_0$, so the constituent quark masses $m_i^*$
are independent of $M_0$.

This mass-rescaling parameter plays a key role to systematically
control the divergences of the loop integrals. We will show how
the model can produce physical quantities in the ultraviolet
limit.

\section{PHYSICAL QUANTITIES}
\label{sec_phys}
We  present the prescription to calculate meson masses, meson decay
constants and topological susceptibility in the leading order of the
$1/N_c$ expansion. These quantities are derived from four-, two-point
functions and bubble diagrams, respectively.

\subsection{Pion and Kaon masses}
\label{subsec_mpk}
The masses of pion and kaon are obtained by evaluating the poles of
their propagators
\begin{equation}
  \Delta_{\mathrm P} (k^2)
  = \frac{2K_{\alpha} }{1-2K_{\alpha}\Pi_{\mathrm P}(k^2)},
\label{meson_prop}
\end{equation}
using the random-phase approximation and the $1/N_c$ expansion, where
$\alpha$ labels the channel isospin and ${\mathrm P}$ denotes the
meson species. The explicit form of the flavor-dependent effective
couplings $K_{\alpha}$ is given by
\begin{align}
K_3 &\equiv G + \frac{1}{2} K (i\,{\rm tr} S^s) ,
     \quad {\rm for} \quad \pi^0,
\label{K_3}\\
K_6 &\equiv G + \frac{1}{2} K (i\,{\rm tr} S^u) ,
     \quad {\rm for} \quad {\mathrm K}^0,\, \bar{\mathrm K}^0.
\label{K_6}
\end{align}
$\Pi_{\mathrm P}$ is the meson self-energy, 
\begin{align}
  \Pi_{\pi}(k^2) &= 2 \Pi_5^{uu}(k^2), \label{Pipi} \\
  \Pi_{\mathrm K}(k^2) &= 2 \Pi_5^{su}(k^2), \label{PiK}
\end{align}
where $\Pi_5^{ij}(k^2)$ is the following loop integral:
\begin{align}
  \Pi_5^{ij}&(k^2)
  = \int \frac{d^D p}{i(2\pi)^D}{\rm tr}
    \bigl[ \gamma_5 S^i(p+k/2) \gamma_5 S^j(p-k/2) \bigr]
    \nonumber \\
  &= \frac{1}{2}
    \left( \frac{i{\rm tr}S^i}{m_i^*}
          +\frac{i{\rm tr}S^j}{m_j^*} \right)
   +\frac{1}{2}[k^2-(m_i^*-m_j^*)^2] I_{ij}(k^2),
\end{align}
\begin{align}
  I_{ij}(k^2) = \int \! \frac{d^D p}{i(2\pi)^D} \,
   \frac{{\rm tr}1}{ ( p^2-m_i^{* \, 2} ) 
   \bigl[ (p-k)^2-m_j^{*\,2} \bigr] }.
\label{int_Iij}
\end{align}
The trace runs over color and spinor indices, then
${\rm tr}1=2^{D/2} N_c$.

The conditions which determine the pion and kaon masses are
\begin{align}
 & 1-2K_{3} \Pi_{\pi}(m_\pi^2)=0 , \label{pion_eq}\\
 & 1-2K_{6} \Pi_{\mathrm K}(m_{\mathrm K}^2)=0. \label{kaon_eq}
\end{align}
These equations are the relations between model parameters and
the input physical quantities $m_{\pi}$ and $m_{\mathrm K}$.

\subsection{$\eta$ and $\eta^{\prime}$ masses}  
$\eta$ and $\eta'$ mesons
are the mass eigenstates for $\eta_8$ and $\eta_0$ mixing. In the
random-phase approximation and the $1/N_c$ expansion, the propagator
of the $\eta-\eta'$ system is given
by~\cite{Klevansky:1992qe,Hatsuda:1994pi}
\begin{equation}
  {\boldsymbol \Delta}^+(k^2)
  = \frac{ 2{\boldsymbol K}^+ }
       { 1-2 {\boldsymbol K}^+ {\boldsymbol \Pi}(k^2)} ,
\label{pro^+}
\end{equation}
where the effective coupling ${\boldsymbol K}^+$ and the self-energy 
${\boldsymbol \Pi}$ are the $2\times2$ matrices
\begin{eqnarray}
{\boldsymbol K}^+ &=& \left(
\begin{array}{cc}
 K_{00} & K_{08} \\
 K_{80} & K_{88}
\end{array}
\right) ,
\label{K^+} \\
{\boldsymbol \Pi} &=& \left(
\begin{array}{cc}
 \Pi_{00} & \Pi_{08} \\
 \Pi_{80} & \Pi_{88}
\end{array}
\right) ,
\label{Pi}
\end{eqnarray}
with
\begin{align*}
&K_{00}=G - \frac{1}{3} K(i\,{\rm tr} S^s + 2i \,{\rm tr}S^u),
\\ 
&K_{88}=G - \frac{1}{6} K(i\,{\rm tr} S^s - 4i\,{\rm tr}S^u),
\\ 
&K_{08}=K_{80}=- \frac{\sqrt{2}}{6} 
 K(i\,{\rm tr} S^s - i\,{\rm tr}S^u),
\end{align*}
and
\begin{align*}
&\Pi_{00}(k^2)=\frac{2}{3}\left[ 2\Pi_5^{uu}(k^2)+\Pi_5^{ss}(k^2) \right],
\\ 
&\Pi_{88}(k^2)=\frac{2}{3}\left[ \Pi_5^{uu}(k^2)+2\Pi_5^{ss}(k^2) \right],
\\ 
&\Pi_{08}(k^2)=\Pi_{80}(k^2)=\frac{2\sqrt{2}}{3} \left[ \Pi_5^{uu}(k^2)
                 -\Pi_5^{ss}(k^2) \right].
\end{align*}

The masses of $\eta$ and $\eta'$ are obtained by solving
\begin{equation}
 {\rm det}[1-2 {\boldsymbol K}^+ 
 {\boldsymbol \Pi}(m_{\mathrm P}^2)] = 0,
\label{eta-eta'}
\end{equation}
where $m_{\mathrm P}$ denotes the mass of $\eta$ or $\eta^{\prime}$.
They are also obtained via diagonalization of Eq.~(\ref{pro^+}) 
~\cite{Hatsuda:1994pi, Rehberg:1995kh}.

\subsection{Pion and Kaon decay constants}  
The decay constants of pion and kaon,  $f_{\rm P}$ are defined by the
matrix element of axial current between the meson and vacuum states,
\begin{align}
 &ik_\mu f_{\mathrm P} \delta_{\alpha\beta} \nonumber \\
 &= - M_0^{4-D} \!\int \! \frac{d^D p}{(2\pi)^D}
  {\rm tr}\left[\gamma_\mu \gamma_5 \frac{T_\alpha}{2} 
  S^i g_{{\mathrm P}qq} \gamma_5 T_\beta^\dagger S^j \right],
\end{align}
where the meson-quark-quark effective coupling $g_{{\mathrm P}qq}$
is defined by
\begin{equation}
  {(g_{{\mathrm P}qq})}^{-2} = M_0^{4-D}
  \frac{\partial\Pi_{\mathrm P}(k^2)}{\partial k^2}.
\label{g_Pqq}
\end{equation}
For notational simplicity, when it is obvious, we will omit the
subscript ``${\rm{rs}}$''  indicating that the quantity is
rescaled with the help of a power of $M_0$. In the leading order
of $1/N_c$ expansion, the decay constants are calculated
as~\cite{Inagaki:2010nb},
\begin{align}
 f_{\pi}^2 &= m_u^{*2} M_0^{4-D} I_{uu}(0) ,
 \label{f_pi} \\
 f_{\mathrm K}^2
 & = \frac{M_0^{4-D}}{J_{us}(0)} \Bigl[m_u^* I_{us}(0) 
   + (m_s^* -m_u^*)
   \nonumber \\
 &\quad \times {\rm tr} \int_0^1 \! dx \int \! \frac{d^D p}{i(2\pi)^D}
  \frac{x}{ \{ p^2 - L_{us}(0) + i\varepsilon \}^2} \biggr]^2 ,
  \label{f_K}
\end{align}
where $J_{us}$ is defined by
\begin{align}
  J_{us}&(k^2) = I_{ij}(k^2) + 2 \{(m_s^* - m_u^*)^2 - k^2 \}
  \nonumber \\
  & \times {\rm tr} \int_0^1 \! dx \int \! \frac{d^D p}{i(2\pi)^D} 
  \frac{x(1-x)}{ \{ p^2 - L_{us}(k^2) + i\varepsilon \}^3} ,
\label{J_us} 
\end{align}
with
\begin{equation}
L_{ij}(k^2) = m_i^{*2} - (m_i^{*2} - m_j^{*2})x - k^2 x(1-x).
\nonumber
\end{equation}
Equation~(\ref{f_pi}) is used to fix the mass scale parameter $M_0$.

\subsection{Topological susceptibility}  
The topological susceptibility 
$\chi$ is defined by the correlation function between the topological 
charge densities, $Q(x)$, at different points~\cite{Hatsuda:1994pi},
\begin{equation}
\chi =\, \int d^D x \langle 0| TQ(x)Q(0) 
 |0\rangle_{\rm connected} ,
\label{chi}
\end{equation}
where
\begin{align}
Q(x)\equiv\frac{g^2}{32\pi^2}F^a_{\mu\nu} \tilde{F}^{a\mu\nu}
= 2K\, {\rm Im} [\det \bar{q}(1-\gamma_5)q] ,
\end{align}
$g$ is the strong coupling constant of QCD and $F^a_{\mu\nu}$ 
is the field strength for gluons. Equation~(\ref{chi}) should be
multiplied by $M_0^{4-D}$ to adjust the mass dimensions. 
In the leading order of $1/N_c$ expansion $\chi$ is given
by~\cite{Fukushima:2001hr}
\begin{align}
 \chi
 =\, &\frac{4K^2}{M_0^{D-4}}(i{\rm tr}S^u)^2
 \left[ (i{\rm tr}S^u)(i{\rm tr} S^s) 
 \left(\frac{2i{\rm tr}S^s}{m_u^*} + 
 \frac{i{\rm tr}S^u}{m_s^*} \right) \right.
 \nonumber \\
 &+\left\{ \frac{1}{\sqrt{6}} (2i{\rm tr}S^s + i{\rm tr}S^u)
 \bigl( \Pi_{00}(0), \Pi_{08}(0) \bigr) \right.
 \nonumber \\
 &+\left. \frac{1}{\sqrt{3}} (i{\rm tr}S^s - i{\rm tr}S^u)
 \bigl( \Pi_{08}(0), \Pi_{88}(0) \bigr)
 \right\} {\boldsymbol \Delta}^+(0)
 \nonumber \\
 &\times\left\{ \frac{1}{\sqrt{6}} (2i{\rm tr}S^s + i{\rm tr}S^u)
 \left(
 \begin{array}{c}
  \Pi_{00}(0) \\
  \Pi_{08}(0) 
 \end{array}\right) \right.
 \nonumber \\
 &+\left. \frac{1}{\sqrt{3}} (i{\rm tr}S^s - i{\rm tr}S^u)
 \left( \left.
 \begin{array}{c}
  \Pi_{08}(0) \\
  \Pi_{88}(0)
 \end{array} \right) \right\} \right] .
\label{chi2}
\end{align}
Thus $\chi$ is evaluated by the quantities already
obtained above.

\section{Strategy}
\label{sec_strategy}
As mentioned in the introduction, our goal is to test the
regularization parameter dependence (or independence) in the
NJL model.

We use the abstract denotations: ${\mathsf m}(\in{\mathcal M})$
are the model predictions, ${\mathsf p}(\in{\mathcal P})$ are the
model parameters, and ${\mathsf i}(\in{\mathcal I})$ are the input
physical quantities. The capitals  ${\mathcal M}, {\mathcal P}$ and
${\mathcal I}$ represent the sets of these quantities. The model
predictions can be performed if all the parameters are known, so
the model determines some function ${\mathcal F}_{\mathsf{mp}}$,
\begin{equation}
  {\mathsf m}^{{\mathrm R}}({\mathcal P})
  = {\mathcal F}_{\mathsf{mp}}^{\mathrm R}
   ({\mathcal P}),
\end{equation}
where the superscript ${\mathrm R}$ indicates a regularization
procedure. In the similar manner one can obtain the functions
${\mathcal F}_{\mathsf{pi}}^{\mathrm R}$ which connect
${\mathcal P}$ and ${\mathcal I}$ through the parameter fitting.
Thus, the model  relates
${\mathcal I}$ to ${\mathcal M}$, as
\begin{equation}
  {\mathsf m}^{\mathrm R}({\mathcal I})
  = {\mathcal F}_{\mathsf{mi}}^{\mathrm R}
   ({\mathcal I}).
\end{equation}
The resulting values of ${\mathsf m}$ should not depend on
regularization methods, if they correctly capture the physics
in question. In the next section, we obtain these functions
${\mathcal F}^{\mathrm R}_{\mathsf{mi}}$,
${\mathcal F}^{\mathrm R}_{\mathsf{mp}}$
and ${\mathcal F}^{\mathrm R}_{\mathsf{pi}}$.

Our model with the DR has six free parameters
${\mathcal P}_6=\{m_u,m_s,G,K,D,M_0\}$. Four of them,
${\mathcal P}_4=\{m_s,G,K,M_0\}$, may be fitted  to the input
meson properties,
${\mathcal I}_4=\{m_{\pi},f_{\pi},m_{\mathrm K},m_{\eta^{\prime}}\}$.
After this partial parameter setting, predicted meson properties,
${\mathcal M}= \{f_{\mathrm K}, m_{\eta}, \chi, \dots \}$ are
written as the functions of the remaining parameters $m_u$ and $D$,
\begin{equation}
  {\mathsf m}^{\rm DR}(m_u,D)
  = {\mathcal F}_{\mathsf{mp}}^{\rm DR}
  (m_u,D)|_{{\mathcal P}_4 \leftarrow {\mathcal I}_4}.
\end{equation}
We cannot obtain finite values for all predicted quantities in the
4D limit, all of ${\mathsf m}$ are finite at $D<4$. However, it is
possible to obtain finite values for some of ${\mathsf m}$ at the
four dimensional limit. Furthermore, some of physical quantities
may be independent of the parameter $D$~\cite{Inagaki:2010nb}.
We are interested in this aspect of the NJL model  which does not
depend on the regularization parameter.  Namely, it is interesting
that in the model  discussed here there are theoretical relations 
between some observables which are finite  (although the model is
not renormalizable)  and, moreover, they do not depend on the
regularization parameter. We mean the relations between input and
predicted observables derived with the help of the fitting parameters
procedure.

Once we put the actual numbers into ${\mathcal I}_4$ and fix $D$,
${\mathsf m}$ become the functions of $m_u$. In particular, we shall
evaluate them in the $D \to 4$ limit, then compare between each other
results obtained in the three regularization ways: (1) DR with
$D \to 4$, (2) DR with fixed $D$ by $\chi$, and (3) three-momentum
(${\rm 3M}$) cutoff method. These regularization schemes can
symbolically be written as
\begin{align}
 &{\rm (1)}\, {\mathsf m}^{{\rm 4D}}(m_u)
 ={\mathcal F}_{\mathsf{mp}}^{\rm DR}
  (m_u)|_{{\mathcal P}_4 \leftarrow {\mathcal I}_4, D\to 4}
  ,\nonumber \\
 &{\rm (2)}\, {\mathsf m}^{{\rm DR}}(m_u)
 ={\mathcal F}_{\mathsf{mp}}^{\rm DR}
  (m_u)|_{\{{\mathcal P}_4,D\} \leftarrow \{{\mathcal I}_4,\chi\}}
  ,\nonumber \\
 &{\rm (3)}\, {\mathsf m}^{{\rm 3M}}(m_u) 
 ={\mathcal F}_{\mathsf{mp}}^{\rm 3M}
  (m_u)|_{ {\mathcal P}_4^{\rm 3M} \leftarrow {\mathcal I}_4 },
\nonumber
\end{align}
where ${\mathcal P}^{\rm 3M}_4=\{m_s,G,K,\Lambda \}$, with the
momentum cutoff $\Lambda$. Note that the total number of the
parameters in the cutoff method is five~\cite{Inagaki:2010nb},
since this method does not need the mass parameter.

\section{Four dimensional limit}
\label{sec_4D}
We demonstrate the calculation in the four dimensional (4D) limit
and express the predicted physical quantities in terms of
the input meson properties. By virtue of taking the 4D limit, the
relations among the physical quantities can be simplified through
the leading order of $\epsilon$ expansion. It is even possible to
obtain analytic expressions for some quantities, which will be
discussed below.

\subsection{The summary of the Sec.~\ref{sec_phys}}
To make the model prediction we need to know the following six
quantities:
\begin{equation}
  m_u^*,\, m_s^*,\, G,\, K,\, m_s,\, M_0.
\label{para}
\end{equation}
These are evaluated by the gap equations
\begin{align}
 &m_u^{*} =
  m_u + 4 G (i\, {\rm tr} S^u) + 2 K (i\, {\rm tr} S^u)
  (i\, {\rm tr} S^s),
   \label{gap_u} \\
 &m_s^{*} =
  m_s + 4 G (i\, {\rm tr} S^s) + 2 K (i\, {\rm tr} S^u)^2,
   \label{gap_s}
\end{align}
and the conditions for meson properties
\begin{align}   
 & 1-2K_{3} \Pi_{\pi}(m_\pi^2)=0 ,
   \label{C_pion}\\
 & 1-2K_{6} \Pi_{\mathrm K}(m_{\mathrm K}^2)=0,
    \label{C_kaon} \\
 & {\rm det}[1-2 {\boldsymbol K}^+ 
   {\boldsymbol \Pi}(m_{\eta^{\prime}}^2)] = 0,
    \label{C_eta} \\
 & f_{\pi}^2 = m_u^{*2} M_0^{4-D} I_{uu}(0).
    \label{C_f_pi}
\end{align}
In the following, we will calculate the quantities (\ref{para}) by
solving these six equations in the 4D limit.

\subsection{Constituent quark masses $m_u^*$, $m_s^*$}
The constituent quark mass $m_u^*$ is obtained from Eqs.~(\ref{gap_u})
and (\ref{C_pion}). The gap equation~(\ref{gap_u}) can be rewritten as
\begin{align}
  m_u^{*} = m_u + 4 K_3 (i{\rm tr} S^u),
  \label{gap_u2} 
\end{align}
which in combination with Eq.~(\ref{C_pion}) enables one to get rid
of $K_3$  and thereby to calculate $m_u^*$ without knowing the values
of $G$ and $K$.

In the 4D$(\epsilon \to 0)$ limit, the chiral condensate
$i{\rm tr} S^i$ and the self-energy loop integral $\Pi_5^{ij}(k^2)$
can be expanded in powers of $\epsilon$ as
\begin{align}
 &i\, {\rm tr} S^i = -\frac{N_c}{4\pi^2\epsilon} m_i^{*3},
  \label{trSe} \\
 & \Pi_5^{ij}(k^2)   
 = \frac{N_c}{8\pi^2 \epsilon}
    \bigl[ k^2-2(m_i^{*2}+m_j^{*2}-m_i^* m_j^*) \bigr].
  \label{Pie}
\end{align}
After some algebra we arrive at 
\begin{equation}
 m_u^* = -\frac{m_\pi^2}{4m_u}
  \left\{ 1+\sqrt{1+\frac{8m_u^2}{m_\pi^2}} \right\}.
\label{m_u*}
\end{equation}
Here we adopted the negative solution of $m_u^*$, since the positive 
solution is unstable.

We thus find the analytic expression for $m_u^*$ as a function of
$m_{\pi}$. On the other hand we apply numerical analysis to obtain
$m_s^*$ by simultaneously solving Eqs.~(\ref{C_pion}), (\ref{C_kaon})
and (\ref{C_eta}).

\subsection{Couplings $G$, $K$}
The couplings $G$ and $K$ can be written as the functions of
$m_u^*$, $m_s^*$, $m_\pi$ and $m_{\mathrm K}$.

From Eqs.~(\ref{C_pion}) and (\ref{C_kaon}), one derives the
following relations
\begin{align}
 G &= \frac{-i{\rm tr}S^u \Pi_{\mathrm K}
        + i{\rm tr}S^s \Pi_{\pi} }
        {2(i{\rm tr}S^s-i{\rm tr}S^u)\Pi_{\pi}\Pi_{\mathrm K} },
 \label{G_eq}\\
 K &= \frac{\Pi_{\mathrm K} -\Pi_{\pi} }
           {(i{\rm tr}S^s-i{\rm tr}S^u)\Pi_{\pi}\Pi_{\mathrm K} }.
 \label{K_eq}  
\end{align}
where we use the abbreviated notations of $\Pi_{\pi}=\Pi_{\pi}(m_\pi^2)$
and $\Pi_{\mathrm K}=\Pi_{\mathrm K}(m_{\mathrm K}^2)$.
By using Eqs.~(\ref{Pipi}), (\ref{PiK}) and (\ref{Pie})
we arrive at the following expressions for the couplings
\begin{align}
G&= \frac{2\pi^2 \epsilon}{N_c} \frac{1}{m_s^{*3}-m_u^{*3}}
    \left[ \frac{m_s^{*3}}{m_{\mathrm K}^2-2(m_u^{*2}+m_s^{*2}-m_u^* m_s^*)}  
    \right. \nonumber \\
 &\quad \left. - \frac{m_u^{*3}}{m_\pi^2-2 m_u^{*2}} \right]
     +{\mathcal O}(\epsilon^2) ,
 \label{Ge} \\
K&= \frac{16\pi^4 \epsilon^2}{N_c^2} \frac1{m_s^{*3}-m_u^{*3}}
    \left[ \frac1{m_{\mathrm K}^2-2(m_u^{*2}+m_s^{*2}-m_u^* m_s^*)}  
    \right. \nonumber \\
 &\quad \left. - \frac{1}{m_\pi^2-2 m_u^{*2}} \right]
    +{\mathcal O}(\epsilon^3).
 \label{Ke}
\end{align}
Thus the couplings are the
functions of $m_s^*$, $G(m_s^*)$ and $K(m_s^*)$. With the help of
Eq.~(\ref{C_eta}), we can determine $m_s^*$ and substitute it in 
$G(m_s^*)$ and $K(m_s^*)$ to get  $G$ and $K$.

Note that $G$ and $K$ are of the order of $\epsilon$ and $\epsilon^2$,
respectively. In the naive $\epsilon \to 0$ limit $G$ and $K$
tend to $0$~\cite{Wilson:1972cf}.
However, the combinations $G {\rm tr}S$ and $K ({\rm tr}S)^2$
appears to be finite because the chiral condensate, Eq.~(\ref{trSe}),
is of the order of $\epsilon^{-1}$. Then, the dynamical masses,
Eq.~(\ref{gap_eq}), can have larger values than
$m_i$ even in the limit of $\epsilon \to 0$.

From the order counting of $1/\epsilon$ for Eqs.~(\ref{gap_u}),
(\ref{gap_s}) and (\ref{trSe}), we find the following relation,
\begin{equation}
 \left( \frac1{\epsilon} \right)^{L-N_G-2N_K}
 =\left( \frac1{\epsilon} \right)^{1-n/2} ,
\label{count}
\end{equation}
where $L, N_G$ and $N_K$ are the number of quark loops (chiral
condensates), $G$ and $K$, respectively. $n$ is the number of the
external quark lines ($n=2l$, $l=0,1,2,\cdots$, $l$ are the number
of external meson lines). Then $n$ point functions become finite 
except the bubble diagrams.

For instance, the above discussion is manifested in the expressions 
for the pion and kaon propagators
in the limit $D \to 4$,
\begin{equation}
  \Delta_{\rm P}(k^2)
  = -\frac{4\pi^2\epsilon}{N_c} \frac{1}{k^2-m^2_{\rm P}}.
\label{re_prop}
\end{equation}
These are derived through the substitution of the obtained $G$, $K$ and
$\Pi_{\rm P}$ into Eq.~(\ref{meson_prop}).

\subsection{Current strange quark mass $m_s$}
We have seen that $m_u^*$, $m_s^*$, $G$ and $K$ are determined
from Eqs.~(\ref{gap_u}), (\ref{C_pion}), (\ref{C_kaon}) and
(\ref{C_eta}). Substituting
these into Eq.~(\ref{gap_s}), 
it is easy  to evaluate $m_s$. The numerical result
will be shown in Sec.~\ref{result_m_s}.

\subsection{Mass scale parameter $M_0$}
In this subsection, we shall set the remaining parameter $M_0$,
the mass scale, by using the pion decay constant $f_\pi$ in
Eq.~(\ref{C_f_pi}).

In the 4D limit, the integral $I_{uu}$ (Eq.~(\ref{int_Iij})) becomes
\begin{eqnarray}
\lim_{\epsilon \to 0} I_{uu}
&=& \lim_{\epsilon \to 0} \frac{N_c}{(2\pi)^{2-\epsilon}}\Gamma(\epsilon)
\int_0^1 dx \, L_{ij}^{-\epsilon} \nonumber \\
&\simeq& - \lim_{\epsilon \to 0} \frac{N_c}{4\pi^2\epsilon} .
\label{Iuue}
\end{eqnarray}
Hence we obtain the 
mass scale from Eq.~(\ref{f_pi}):
\begin{equation}
  \lim_{\epsilon \to 0} M_0^{2\epsilon} 
  = \lim_{\epsilon \to 0} \frac{4\pi^2\epsilon}{N_c} \frac{f_\pi^2}{m_u^{*2}} .
\label{M0_D4}
\end{equation}
Since $m_u^*$ is the function of $m_\pi$, (\ref{m_u*}), 
the value of $M_0$ is determined by $m_\pi$ and $f_\pi$.
$M_0$  goes to 0 in the limit of $\epsilon \to 0$.

It is worth mentioning that by using Eqs.~(\ref{M0_D4}),
(\ref{f_K}) and (\ref{J_us}), $f_{\mathrm K}$ can be written as
\begin{equation}
   f_{\mathrm K}^2 = f_\pi^2 \frac{(m_u^* + m_s^*)^2}{4m_u^{*2}} ,
\label{f_K_final}
\end{equation}
since $J_{ij}$ coincides with $I_{ij}$ in the leading order
of the $\epsilon$ expansion.

\subsection{Chiral condensates
$\langle \bar{u}u \rangle$, $\langle \bar{s}s \rangle$}
From the Eq.~(\ref{trSe})  the chiral
condensates $\langle \bar{i}i \rangle$ are of the order of $\epsilon^{-1}$, so they
diverge in the $\epsilon \to 0$ limit. However, the 
rescaled chiral condensates $\langle \bar{i}i \rangle_{\rm{rs}}$
($\equiv M_0^{4-D} \langle \bar{i}i \rangle$) are finite,
since the  order of $M_0^{4-D}$ is $\epsilon$ as seen in
Eq.~(\ref{M0_D4}).

The form of the rescaled chiral condensate 
$\langle \bar{u}u \rangle_{\rm{rs}}$ is obtained from
Eqs.~(\ref{trSe}), (\ref{m_u*}) and (\ref{M0_D4}) as
\begin{equation}
\langle \bar{u}u \rangle_{\rm{rs}} = -\frac{m_\pi^2 f_\pi^2}{4m_u} 
\left\{ 1+\sqrt{1+\frac{8m_u^2}{m_\pi^2}} \right\}.
\label{uu}
\end{equation}
If we ignore the last term of Eq.~(\ref{uu}), the above equation
coincides with the Gell-Mann--Oakes--Renner relation~%
\cite{GellMann:1968rz, Glashow:1967rx}.
With the help of Eq.~(\ref{f_K_final}), one has the
analytic expression for $\langle \bar{s}s \rangle_{\rm rs}$:
\begin{equation}
 \langle \bar{s}s \rangle_{\rm{rs}}
 = -\frac{m_\pi^2 f_\pi^2}{4m_u} 
   \left( 2\frac{f_{\mathrm K}}{f_\pi} -1 \right)^3
   \left\{ 1+\sqrt{1+\frac{8m_u^2}{m_\pi^2}} \right\}.
\label{ss}
\end{equation}
In the limit $f_{\mathrm K} \to f_\pi$, $\langle \bar{s}s \rangle_{\rm{rs}}$
coincides with $\langle \bar{u}u \rangle_{\rm{rs}}$.

\section{Numerical results}
\label{result}
To evaluate the physical quantities discussed in Sec.~\ref{sec_phys}
we employ the following input meson properties:
\begin{eqnarray*}
 & m_\pi=138{\rm MeV},\ & f_\pi=92{\rm MeV},\\
 & m_{\mathrm K}=495{\rm MeV},\   & m_{\eta'}=958{\rm MeV}.
\end{eqnarray*}
We first evaluate the constituent quark masses and the strange
current quark mass, then calculate the meson properties
$f_{\mathrm K}$, $m_{\eta}$, and the topological susceptibility
$\chi$.

As mentioned in Sec.~\ref{sec_strategy}, we compare the results obtained
in the the three cases: (1) DR with 4D, (2) DR with $D(\chi)$,
and (3) 3M cutoff method. In the second case, we select the value
$\chi^{1/4}=170$MeV~\cite{Inagaki:2010nb} and 
call it simply the DR results.

\subsection{Constituent quark masses $m_u^*$, $m_s^*$}
\begin{figure}[!h]
  \begin{center}
    \includegraphics[height=2.0in,keepaspectratio]
    {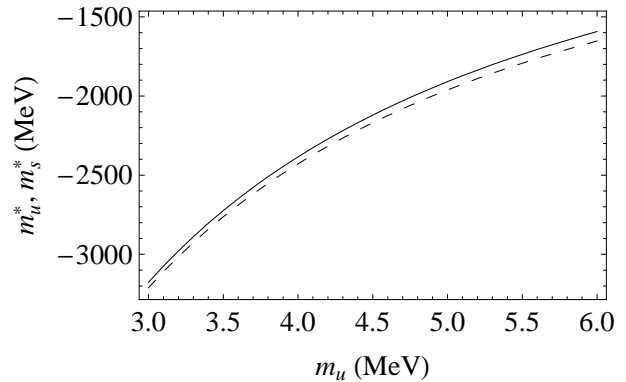} 
  \end{center}
  \vspace{-0.5cm}
  \caption{$m_u^*$ and $m_s^*$
   as the functions of $m_u$. The solid and dashed lines are the
   results of $m_u^*$ and $m_s^*$, respectively.}
  \label{mias}
\end{figure}

%
Before evaluating the observed quantities, we consider an
unobserved quantity, the constituent quark mass.
In Fig.~\ref{mias}, $m_u^*$ and $m_s^*$ are shown as the functions
of $m_u$ in the range $3{\rm MeV} \leq m_u \leq 6{\rm MeV}$.
Since $m_u$ is contained in the denominator of Eq.~(\ref{m_u*}),
 $m_u^*$ depends strongly on $m_u$. 
Note that $m_u^*$ and $m_s^*$ can be positive according the discussion
in Ref.\cite{Inagaki:2012re}.
The values of $|m_u^*|$ and
$|m_s^*|$ are considerably larger than in the frequently
used 3M cutoff case, $m_u^* \sim 300$MeV and $m_s^* \sim 500$MeV. 
Therefore a large regularization dependence is found for an unobserved
quantity. Note that $m_s^*$ can be obtained
by solving Eq.~(\ref{f_K_final}) analytically. However, this
solution does not lead a realistic value of $m_{\eta'}$. In 
other words the observed value of the kaon decay constant, $f_{\mathrm K}$, 
can not be consistent with the realistic value for $m_{\eta'}$.

\subsection{Current quark mass $m_s$}
\label{result_m_s}
\begin{figure}[!h]
  \begin{center}
    \includegraphics[height=2.0in,keepaspectratio]
    {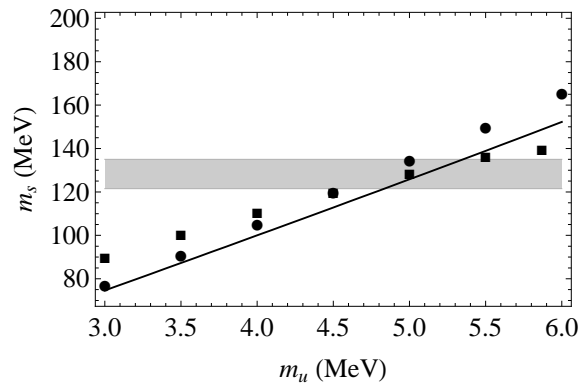} 
  \end{center}
  \vspace{-0.5cm}
  \caption{The strange quark mass $m_s$ as the function of $m_u$.
   The black line shows $m_s$ in the 4D case. The circles and
   squares are DR and cutoff regularization results, respectively.
   The gray bound is the experimental region.}
  \label{ms}
\end{figure}

The plots of $m_s$ obtained in the 4D, DR and cutoff cases
are shown in Fig.~\ref{ms}.
We see that $m_s$ increases linearly with respect to $m_u$.
It is interesting to note that the DR and cutoff results
show behavior similar to the 4D result. 
The results around $m_u \simeq 5$MeV cross
the experimental region which is evaluated at
1GeV~\cite{Nakamura:2010zzi}.

\subsection{Kaon decay constant $f_{\mathrm K}$}
\begin{figure}[!h]
  \begin{center}
    \includegraphics[height=2.0in,keepaspectratio]
    {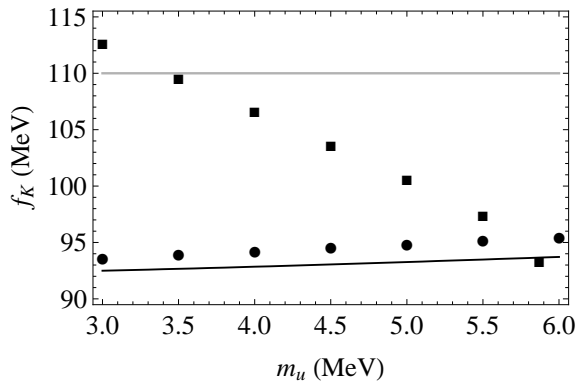} 
  \end{center}
  \vspace{-0.5cm}
  \caption{The kaon decay constant $f_{\mathrm K}$ as the function of
   $m_u$. Black and gray lines are the obtained $f_{\mathrm K}$ and
   the experimental value.
   The circles and squares are DR and cutoff
   regularization results, respectively.}
  \label{fig_fk}
\end{figure}

%
Figure~\ref{fig_fk} displays the results of $f_{\mathrm K}$
with the experimental value $f^{\rm ex}_{\mathrm K}=110$MeV.
The resulting $f_{\mathrm K}$ in the 4D case is smaller than its
experimental value. The DR plots  are similar to the 4D case, and
they are a few MeV closer to the experimental line. On the other
hand, $f_{\mathrm K}$ decreases with increasing $m_u$ in the cutoff
regularization, which is the opposite to the 4D and DR cases
tendency. The cutoff results for small $m_u$ region receive the
large effect of the parameter dependence.

\subsection{Eta meson mass $m_{\eta}$}
\begin{figure}[!h]
  \begin{center}
    \includegraphics[height=2.0in,keepaspectratio]
    {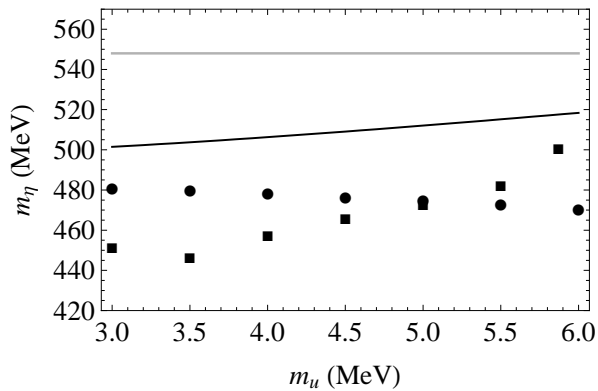} 
  \end{center}
  \vspace{-0.5cm}
  \caption{The eta meson mass $m_\eta$ as the function of $m_u$.
   The black and gray lines are the obtained $m_\eta$ and the
   experimental value, $m_{\eta}^{\rm ex}=548$MeV. The circles
   and squares are the DR and the cutoff results.}
  \label{fig_meta}
\end{figure}

We put the result of $m_\eta$ in Fig.~\ref{fig_meta}.
In the 4D case, $m_{\eta}$ is around $500$MeV at $m_u=3.0$MeV, and
it slightly increases with respect to $m_u$. For all the region,
the values are smaller than in the experimental data. Contrary to the
results seen in $f_{\mathrm K}$, both the DR and cutoff cases are
worse than the 4D case in terms of comparison with the experimental data.

\subsection{Topological susceptibility $\chi$}
\begin{figure}[!h]
  \begin{center}
    \includegraphics[height=2.0in,keepaspectratio]{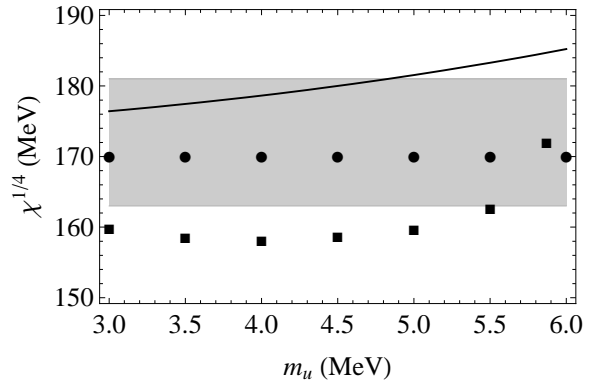} 
  \end{center}
  \vspace{-0.5cm}
  \caption{The topological susceptibility $\chi$ as the
   function of $m_u$. The black line is the obtained $\chi$
   in the 4D. The circles and squares are the DR and cutoff results,
   respectively. The gray bound is the result of lattice simulation~%
   \cite{Alles:1996mn}.} 
\label{fig_chi}
\end{figure}

%
In Fig.~\ref{fig_chi}, $\chi$ is calculated as a function
of $m_u$.
One sees that, in the 4D case, $\chi$ becomes larger when $m_u$
increases. The DR result is trivially fixed at
$\chi^{1/4}=170$MeV, because $\chi$ is the fifth
input quantity in this case. The gray bound shows the results of
lattice simulation~\cite{Alles:1996mn}, $\chi^{1/4}=170\pm7,
174\pm7$MeV, and Witten-Veneziano mass formula~\cite{Witt:79,Vene:79},
$\chi^{1/4}=179$MeV. The 4D result is close to the values
of lattice simulation. The 4D and DR results are plotted inside the
lattice region. However, almost all squares are located outside of
the lattice region, so the cutoff case is worse than the 4D and DR
in this context.

\subsection{Chiral condensates 
  $\langle \bar{u}u \rangle$, $\langle \bar{s}s \rangle$}

\begin{figure}[!h]
  \begin{center}
    \includegraphics[height=2.0in,keepaspectratio]{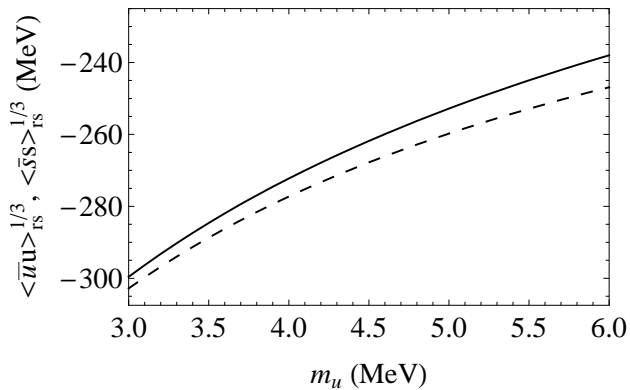} 
  \end{center}
  \vspace{-0.5cm}
  \caption{The chiral condensates $\langle \bar{u}u \rangle_{\rm{rs}}$
   and $\langle \bar{s}s \rangle_{\rm{rs}}$ as the function of $m_u$.
   The solid and dashed lines 
   indicate $\langle \bar{u}u \rangle_{\rm{rs}}$ and
   $\langle \bar{s}s \rangle_{\rm{rs}}$, respectively.}
\label{fig_ii}
\end{figure}

The chiral condensates are shown as functions of $m_u$ in
Fig.~\ref{fig_ii} where we use the obtained $f_{\mathrm K}$ to
evaluate Eq.~(\ref{ss}).
The $m_u$ dependence can easily be read off the explicit forms
in Eqs.~(\ref{uu}) and (\ref{ss}).
It is interesting to note that, although the values of
$|m_i^*|$ in the 4D and DR methods are much larger
than in the cutoff case, $m_i^* \sim \mathcal{O}(100{\rm MeV})$,
the chiral condensates almost coincide in the 4D and DR regularizations.

\section{Conclusion}
\label{conclusion}
We have constructed a NJL model with DR which is convergent in
the 4D limit, and found that it is possible to control the ultraviolet
divergences within the framework mentioned in the section~\ref{sec_4D}.
The obtained results describe meson properties pretty well. We believe
that the treatment prescribed here has possibilities for future work.

(I) We have shown that the model is free of
divergences by virtue of introducing the mass scale parameter.
The mass scale runs according  the dimension in the dimensional
regularization scheme where the ultraviolet limit corresponds to
4D limit. We find that it is possible to remove all the
divergence within the model frame work. Here the mass-scaling
determines the scale of the model, so this is reminiscent
of the renormalization flow. We have explicitly shown the
ultraviolet behavior via numerical calculations.

(II) 
It is a non-trivial question whether so constructed model can
produce reliable results, because important contributions may be
dropped due to taking the 4D limit. However the obtained results
indicate that the model behaves well even in this limit.

(III)
The meson properties are described without the regularization
parameter. The model predictions do not depend on $D$ due to
the effect of the mass rescaling. This indicates that we can
consider the ultraviolet limit even in the nonrenormalizable
model.

(IV) We found that, in this model framework, it is possible
to obtain analytical relations between meson properties and
the chiral condensates. This point has practical importance
as the model calculations are significantly simplified. In
particular, Eq.~(\ref{re_prop}) has quite simple form for the
meson propagators. This form implies that we can introduce
meson propagators in the same form as it appears in the usual
perturbative expansion techniques. Therefore the method is
expected to be useful in more complicated problems such as 
for example in the three-body formalism.

We have also clarified the following less important points:

(V) By applying the effects of the mass scaling, we have clarified
the role of the mass dimension. It is known that the couplings
should become zero in the ultraviolet limit so that the theory
well behaves in the ultraviolet limit~\cite{Eguchi:1977kh}.
We have effectively incorporated the ``running'' couplings by
using the mass rescaling to control the divergences with keeping
the finite values of the rescaled couplings.

(VI) Related to the point (I) we have introduced the effective
coupling scale in the couplings $G$ and $K$, and this works successfully. 
This is trivial if one deals with the renormalized theory where
the ultraviolet behavior is well known through the renormalization
group flow. However, in the NJL model, the background gluon
contributions are implicitly expressed in terms of the effective coupling
strength. Therefore the couplings 
regularized in specific ways contain important dynamical
information of the model. Our results indicate that the model with
the DR does not miss dominant physical contributions when we
take the high-energy limit.

The model predictions are intimately related to the employed
regularization prescriptions, because they include the background
dynamical information as was explained in the point (VI).
Then, it may be interesting to investigate whether
the ultraviolet behavior of the model
different regularization schemes leads to the similar consequences.
We believe the DR method is especially good in this context,
because it is expected to preserve the required symmetry of the
model just as it does quite successfully in the formal perturbative
quantum field theories. This statement is confirmed by
phenomenological results of the paper.

Concerning the point (II) we also found that the physical quantities
calculated in the DR with $D<4$ and in the cutoff regularization 
have values similar to those obtained in the DR at the 4D limit.

However, there is a discrepancy between the model results and the
experimentally observed values. We believe that this discrepancy should
be explained by introducing higher dimensional operators.

Calculations are drastically simplified if one uses our method thanks
to taking the 4D limit. Thus we hope that the regularization parameter
independent approach discussed in this paper can be useful to
systematically introduce higher dimensional operators and to calculate
higher order corrections.

\begin{acknowledgments}
The authors would like to thank Y. Hoshino, T. Morozumi and
K. Ishikawa for fruitful discussions. HK is supported by the
National Research Foundation of Korea funded by the Korean
Government (Grant No. NRF-2011-220-C00011).
AK is supported by the Georgian Shota Rustaveli National Science
Foundation (grant 11/31).
\end{acknowledgments}



\begin{thebibliography}{00}
\bibitem{NJL}
  Y.~Nambu and G.~Jona-Lasinio, Phys. Rev. \textbf{122}, 345 (1961);
  \textbf{124}, 246 (1961).

\bibitem{Vogl:1991qt}
  U.~Vogl and W.~Weise, 
  Prog. Part. Nucl. Phys. \textbf{27}, 195 (1991)

\bibitem{Klevansky:1992qe}
  S.~P.~Klevansky,
  Rev. Mod. Phys.  \textbf{64}, 649 (1992).

\bibitem{Hatsuda:1994pi}
  T.~Hatsuda and T.~Kunihiro,
  Phys. Rept. \textbf{247}, 221 (1994).

\bibitem{Gross:1974jv} 
  D.~J.~Gross and A.~Neveu,
  Phys.\ Rev.\ D {\bf 10}, 3235 (1974).

\bibitem{Eguchi:1977kh} 
  T.~Eguchi,
  Phys.\ Rev.\ D {\bf 17}, 611 (1978).

\bibitem{Shizuya:1979bv} 
  K.~-i.~Shizuya,
  Phys.\ Rev.\ D {\bf 21}, 2327 (1980).

\bibitem{Rosenstein:1988pt} 
  B.~Rosenstein, B.~J.~Warr and S.~H.~Park,
  Phys.\ Rev.\ Lett.\  {\bf 62}, 1433 (1989).

\bibitem{Plant:1997jr}
  R.S.~Plant and M.C.~Birse,
  Nucl. Phys. A \textbf{628}, 607 (1998); 
  A \textbf{703}, 717 (2002).

\bibitem{Hell:2008cc}
  T.~Hell, S.~Roessner, M.~Cristoforetti and W.~Weise,
  Phys. Rev. D \textbf{79}, 014022 (2009); 
  D \textbf{81}, 074034 (2010). 

\bibitem{Krewald:1991tz}
  S.~Krewald and K.~Nakayama, 
  Ann. Phys. \textbf{216}, 201 (1992).

\bibitem{Inagaki:1994ec}
  T.~Inagaki, T.~Kouno and T.~Muta, 
  Int. J. Mod. Phys. A \textbf{10}, 2241 (1995).

\bibitem{Jafarov:2004jw}
  R.G.~Jafarov, and V.E.~Rochev
  Russ. Phys. J. \textbf{49}, 712 (2006).

\bibitem{Inagaki:2007dq}
  T.~Inagaki and D.~Kimura and A.~Kvinikhidze,
  Phys. Rev. D \textbf{77}, 116004 (2008). 

\bibitem{Fujihara:2008ae}
  T.~Fujihara, D.~Kimura, T.~Inagaki and A.~Kvinikhidze
  Phys. Rev. D \textbf{79}, 096008 (2009).

\bibitem{Inagaki:2010nb} 
  T.~Inagaki, D.~Kimura, H.~Kohyama and A.~Kvinikhidze,
  Phys.\ Rev.\ D {\bf 83}, 034005 (2011).

\bibitem{Inagaki:2011uj} 
  T.~Inagaki, D.~Kimura, H.~Kohyama and A.~Kvinikhidze,
  Phys.\ Rev.\ D {\bf 85}, 076002 (2012).

\bibitem{Inagaki:2012re} 
  T.~Inagaki, D.~Kimura, H.~Kohyama and A.~Kvinikhidze,
  Phys.\ Rev.\ D {\bf 86}, 116013 (2012).

\bibitem{Osipov:2004mn}
  A.A.~Osipov, A.H.~Blin and B.~Hiller,
  arXiv:hep-ph/0410148;
  A.A.~Osipov, H.~Hansen and B.~Hiller, 
  Nucl. Phys. A \textbf{745}, 81 (2004).

\bibitem{Inagaki:1997nv}
  T.~Inagaki, S.D.~ Odintsov and Yu.I~ Shil'nov,
  Int. J. Mod. Phys. A \textbf{14}, 481 (1999).

\bibitem{Inagaki:2003yi}
  T.~Inagaki, D.~Kimura and T.~Murata,
  Prog. Theor. Phys. \textbf{111}, 371 (2004).

\bibitem{Kobayashi:1970ji}
  M.~Kobayashi and T.~Maskawa, 
  Prog. Theor. Phys. \textbf{44}, 1422 (1970);
  M.~Kobayashi, H.~Kondo and T.~Maskawa, 
  Prog. Theor. Phys. \textbf{45}, 1955 (1971).

\bibitem{'tHooft:1976fv}
  G.~'t Hooft, 
  Phys. Rev. D \textbf{14}, 3432 (1976);
  \textbf{18}, 2199(E) (1978);
  Phys. Rept. \textbf{142}, 357 (1986).

\bibitem{Wilson:1972cf} 
  K.~G.~Wilson,
  Phys.\ Rev.\ D {\bf 7}, 2911 (1973).

\bibitem{Rehberg:1995kh}
  P.~Rehberg, S.~P.~Klevansky and J.~Hufner,
  Phys.\ Rev.\ C {\bf 53}, 410 (1996).

\bibitem{Fukushima:2001hr}
  K.~Fukushima, K.~Ohnishi and K.~Ohta,
  Phys.\ Rev.\ C {\bf 63}, 045203 (2001).

\bibitem{Nakamura:2010zzi} 
 J. Beringer et al. [Particle Data Group Collaboration],
 Phys.\ Rev.\ D {\bf 86}, 010001 (2012). 


\bibitem{GellMann:1968rz} 
  M.~Gell-Mann, R.~J.~Oakes and B.~Renner,
  Phys.\ Rev.\  {\bf 175}, 2195 (1968).

\bibitem{Glashow:1967rx} 
  S.~L.~Glashow and S.~Weinberg,
  Phys.\ Rev.\ Lett.\  {\bf 20}, 224 (1968).

\bibitem{Alles:1996mn}
  B.~Alles, M.~D'Elia and A.~Di Giacomo,
  Nucl. Phys. B \textbf{494}, 281 (1997);
  B \textbf{679}, 397(E) (2004). 
  
\bibitem{Witt:79}
  E.~Witten, Nucl. Phys. B \textbf{156}, 269 (1979).

\bibitem{Vene:79}
  G.~Veneziano, Nucl. Phys. B \textbf{159}, 213 (1979).


\end{thebibliography}
\end{document}